\documentclass[twocolumn,superscriptaddress,amsmath,amssymb,aps,prl]{revtex4-1}

\usepackage{graphicx}
\usepackage{dcolumn}
\usepackage{bm}

\usepackage[colorlinks,urlcolor=blue,citecolor=blue,linkcolor=blue]{hyperref}

\usepackage{comment}

\newcommand{\vect}[1]{{\mathbf #1}}
    
\newcommand{\up}{\uparrow}
\newcommand{\down}{\downarrow}
\renewcommand{\k}{{\bf k}}

\newcommand{\q}{{\bf q}}
\newcommand{\Q}{{\bf Q}}

\newcommand{\0}{{\bf 0}}

\newcommand{\bra}[1]{\left\langle{#1}\right|}
\newcommand{\ket}[1]{\left|{#1}\right>}

\newcommand{\eb}{E_{\rm B}}

\newcommand{\nn}{\nonumber}
\newcommand{\beq}{\begin{equation}}
\newcommand{\eeq}{\end{equation}}

\newcommand{\mx}{m_{\text{X}}}
\newcommand{\mc}{m_{\text{C}}}
\newcommand{\OR}{\Omega_{\text{R}}}

\newcommand{\exciton}{\hat{b}}

\usepackage[usenames,dvipsnames]{color}
\newcommand{\sout}[1]{}

\begin{document}

\title{Spectroscopic probes of quantum many-body correlations in polariton microcavities}

\author{Jesper Levinsen}
\affiliation{School of Physics and Astronomy, Monash University, Victoria 3800, Australia}

\author{Francesca Maria Marchetti}
\affiliation{Departamento de F\'isica Te\'orica de la Materia Condensada \& Condensed Matter Physics Center (IFIMAC), Universidad Aut\'onoma de Madrid, Madrid 28049, Spain}

\author{Jonathan Keeling}
\affiliation{SUPA, School of Physics and Astronomy, University of St Andrews, St Andrews, KY16 9SS, United Kingdom}

\author{Meera M.~Parish}
\affiliation{School of Physics and Astronomy, Monash University, Victoria 3800, Australia}

\date{\today}

\begin{abstract}
We investigate the many-body states of exciton-polaritons that can be observed by pump-probe spectroscopy. Here, a weak-probe ``spin-down'' polariton is introduced into a coherent state of ``spin-up'' polaritons created by a strong pump.  We show that the $\down$ impurities become dressed by excitations of the $\up$ medium, and form new polaronic quasiparticles that feature two-point and three-point many-body quantum correlations, which, in the low density regime, arise from coupling to the vacuum biexciton and triexciton states respectively. In particular, we find that these correlations generate additional branches and avoided crossings in the $\down$ optical transmission spectrum that have a characteristic dependence on the $\up$-polariton density.  Our results thus demonstrate a way to directly observe correlated many-body states in an exciton-polariton system that go beyond classical mean-field theories.  \end{abstract}

\maketitle

While the existence of Bose-Einstein statistics is fundamentally quantum, many of the properties of Bose-Einstein condensates can be understood from the phenomenology of nonlinear classical waves (see, e.g., Ref.~\cite{Proukakis2013}). In particular, the physics of a weakly interacting gas at low temperatures can generally be described by mean-field theories, involving coherent (i.e., semiclassical) states.  Exceptions to this arise when the strength of interactions becomes comparable to the kinetic energy of the bosons. Here, one has correlated states and even quantum phase transitions, e.g., between superfluid and Mott insulating phases~\cite{Greiner2002a,Bloch2008a}.  For condensates comprised of short-lived bosonic particles such as magnons~\cite{Demokritov2006a}, photons~\cite{Klaers2010}, and exciton-polaritons (superpositions of excitons and cavity photons)~\cite{Kasprzak2006}, the possibility of realizing correlated states suffers a further restriction: the interaction energy scale must exceed the lifetime broadening of the system's quasiparticles.  For these reasons, observing quantum correlated behaviour with such quasiparticles remains a challenging goal. In the case of exciton-polaritons, there has been recent progress in achieving anti-bunching in emission from fully confined photonic dots~\cite{munoz2017:quantum,Delteil2018}.  However, there is ongoing controversy over the strength of the polariton-polariton interaction~\cite{ciuti98:interaction,rochat00:prb,Sun2017}, and there is as yet little known about many-body correlated polariton states.

In this Letter, we propose to engineer and probe quantum correlations in a many-body polariton system through \textit{quantum impurity} physics. Here, a mobile impurity is dressed by excitations of a quantum-mechanical medium, thus forming a new quasiparticle or polaronic state~\cite{mahan2013:many,Devreese2009} that typically defies a mean-field description.  Quantum impurity problems have been studied extensively with cold atoms, where one can explore both Bose~\cite{Hu2016,Jorgensen2016,Camargo2017} and Fermi~\cite{Schirotzek2009,Nascimbene2009,Kohstall2012,Koschorreck2012,Cetina2015doi,Cetina2016,Scazza2018} polarons (corresponding to bosonic and fermionic mediums, respectively).  These studies have yielded insight into the formation dynamics of quasiparticles~\cite{Cetina2016,Parish2016,Shchadilova2016}, and the impact of few-body bound states on the many-body system~\cite{Massignan_Zaccanti_Bruun,yoshida2018}.  Furthermore, in the solid-state context, the Fermi-polaron picture has recently led to a better understanding of excitons immersed in an electron gas~\cite{Sidler2017,Efimkin2017}, as well as the relation of this to the Fermi-edge singularity~\cite{Pimenov2017}.

\begin{figure}[t] \centering \includegraphics[width=\columnwidth]{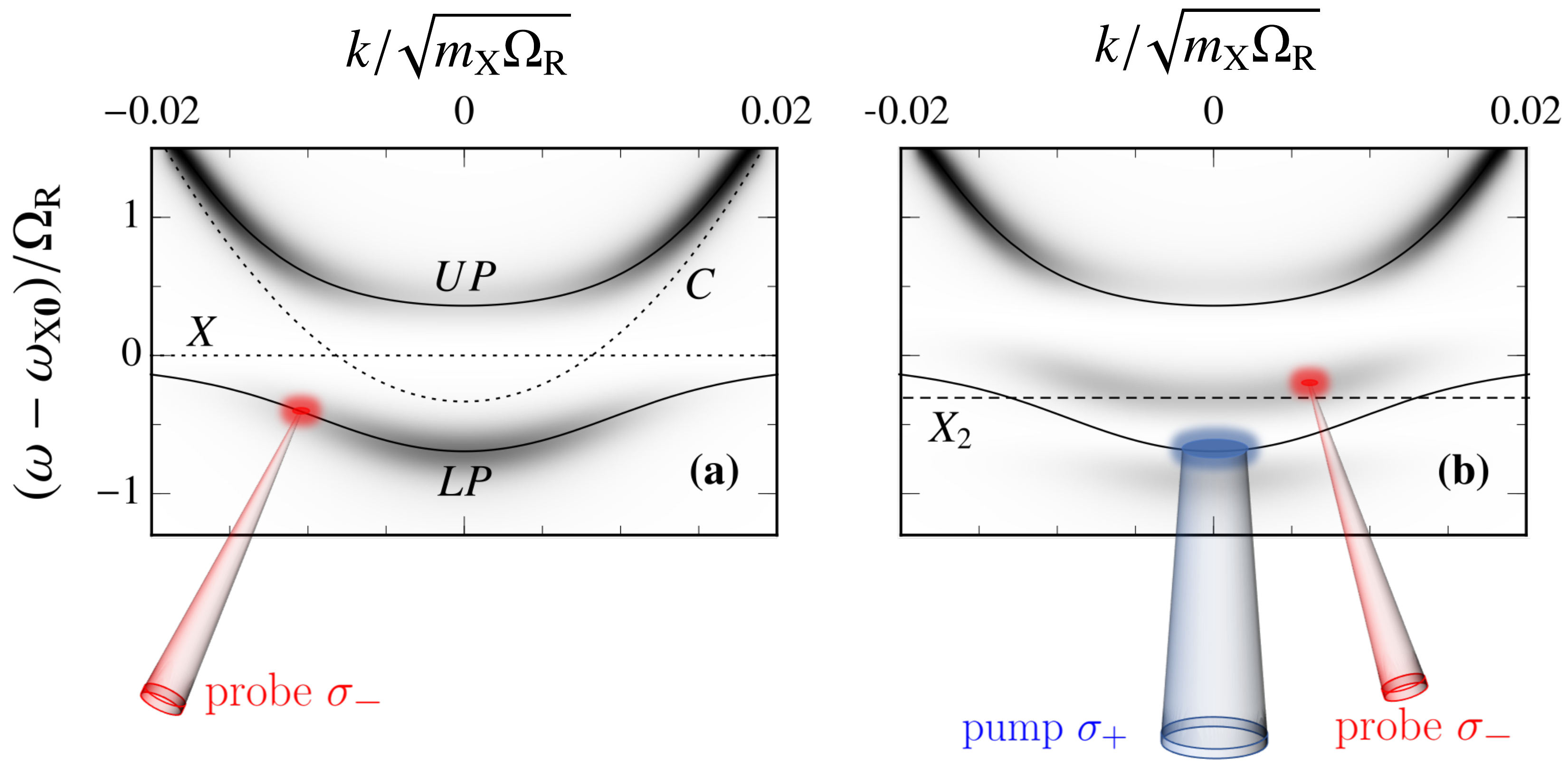} 
\caption{Spectroscopic signature of a two-point many-body correlated state in the probe photon transmission ${\cal T}(\k,\omega)$ (see text and~\cite{supmat}) as a function of momentum and energy.  (a) In the absence of pumping. The upper (UP) and lower (LP) polaritons are shown as solid lines, while the dotted lines correspond to the bare photon (C) and exciton (X) dispersions.  (b) With a $\sigma^+$ pump resonant with the LP at zero momentum.  Resonant coupling to a biexciton ($X_2$) at $\omega+\omega_{\text{LP}\0}\simeq -\eb$ (dashed line) causes a splitting of the bare lower polariton into attractive and repulsive branches, as well as a blue-shift of the upper polariton.  For this illustration, we take the $\sigma_+$ polariton density $n=\mx\OR/8\pi$, detuning $\delta=-\OR/3$, and $\eb=\OR$.  }
\label{fig:dispersion}
\end{figure}

Here we will investigate correlated states of exciton-polaritons using the Bose polaron, which is naturally realized by macroscopically pumping a polariton state in a given circular polarisation ($\up$), and then applying a weak probe of the opposite ($\down$) species (Fig.~\ref{fig:dispersion}).  Indeed, experimental groups have already carried out polarization-resolved pump-probe spectroscopy in the transmission configuration~\cite{Takemura2014,Takemura2017}. However, such measurements were interpreted in terms of a mean-field coupled-channel model involving the vacuum biexciton state~\cite{wouters2007a}, which neglects the possibility of correlated polaronic states.  In particular, there was no analysis of multi-point quantum correlations or how the character of the many-body polaronic state depends on density.

To model the quantum-impurity scenario, we go beyond mean-field theory and construct impurity $\down$-polariton wave functions that include two- and three-point quantum many-body correlations. Such strong multi-point correlations can be continuously connected to the existence of multi-body bound states in vacuum, namely $\down\up$ biexcitons and $\down\up\up$ triexcitons~\cite{Turner2010a} (higher-order bound states have not been observed, as far as we are aware). We calculate the $\down$ linear transmission probe spectrum following resonant pumping of $\up$ lower polaritons, as illustrated in Fig.~\ref{fig:dispersion}(b), and we expose how multi-point correlations emerge as additional splittings in the spectrum with increasing pump strength.  There is thus the prospect of directly accessing polariton correlations from simple spectroscopic measurements performed in standard cryogenic experiments on GaAs-based structures~\cite{Takemura2014,Takemura2017}, as well as in two-dimensional materials at room temperature~\cite{flatten2016room,lundt2016room}.

\paragraph*{Model.--} We consider a spin-$\down$ impurity excited by a $\sigma_{-}$ probe immersed in a gas of spin-$\up$ lower polaritons excited by a $\sigma_+$ pump (see schematic in Fig.~\ref{fig:dispersion}).  The $\down$ probe is optical, but the coupling between $\up$ and $\down$ polarizations arises through the excitonic component.  To capture the effect of the medium on this photonic component, it is natural to describe the impurity in terms of excitons ($\exciton_{\k}^{}$), with dispersion $\omega_{\text{X}\k} = \frac{\k^2}{2\mx}$, and photons ($\hat{c}_{\k}^{}$) with dispersion $\omega_{\text{C}\k} = \frac{\k^2}{2\mc}+\delta$.  Here $\delta$ is the photon-exciton detuning (we take $\omega_{\rm{X}\0}=0$), $\mx$ is exciton mass and $\mc$ is the photon mass --- in this Letter we always take $\mc/\mx\simeq10^{-4}$.  The photon-exciton coupling of strength $\OR$ leads to the formation of lower (LP) and upper (UP) exciton-polaritons~\cite{hopfield58,pekar58} with dispersion:
\begin{align}
    \omega_{{\scriptsize  \begin{matrix} \text{LP}\\\text{UP}\end{matrix}}\k} & = \frac12\left[\omega_{\text{X}\k}+\omega_{\text{C}\k} \mp \sqrt{(\omega_{\text{C}\k}-\omega_{\text{X}\k})^2+\OR^2}\right] .
\end{align}

We choose a pump that is resonant with the lower polaritons at zero momentum, yielding a macroscopically occupied single-particle $\k=0$ state.  Thus, we use the following Hamiltonian~\cite{supmat} (setting $\hbar$ and the area to 1):
\begin{align} 
\nn
  &\hat{H}  = \sum_{\k}
     \!
  \left[\omega_{\text{X}\k} \exciton_{\k}^\dag
  \exciton_{\k}+
    \omega_{\text{C}\k} \hat{c}_{\k}^\dag
  \hat{c}_{\k}
  +\frac{\OR}{2}\left(\exciton_{\k}^\dag
    \hat{c}_{\k}^{} + \text{h.c.}\right) \right]\\
\nn
  &+ \sum_{\k}\left(\omega_{\text{LP}\k} - \omega_{{\rm LP}\0}\right) \hat{L}_{\k}^\dag
  \hat{L}_{\k}^{}
  \!
  + \sum_{\k,
    \k', \vect{q}} g_{\k\k'} 
\hat{L}_{\k}^\dag
\exciton_{\q-\k}^\dag \exciton_{\q-\k'}^{}
\hat{L}_{\k'}^{} \\
&        + \sqrt{n}\sum_{\k, \vect{q}} 
    g_{\k\vect{0}} 
      \, \exciton_{\q-\k}^\dag \exciton_{\q}^{} \left(\hat{L}_{\k}^\dag 
     +  \hat{L}_{-\k}^{} \right) ,
\label{eq:hamil}
\end{align}
which is measured with respect to the energy of the Bose medium in the absence of excitations, $\omega_{\text{LP}\0}n$, where $n$ is the medium density.  Since only the $\up$ LP mode is occupied, we simplify our calculations by writing the medium in the polariton basis, with the finite-momentum LP creation operator $\hat{L}_{\k}^{\dag}$ and excitation energy $\omega_{\text{LP}\k} - \omega_{{\rm LP}\0}$.  For simplicity, we have assumed that the polariton splitting and detuning are independent of polarization; however it is straightforward to generalize our results to polarization-dependent parameters.

\begin{figure*}[t] \centering \includegraphics[width=2.05\columnwidth]{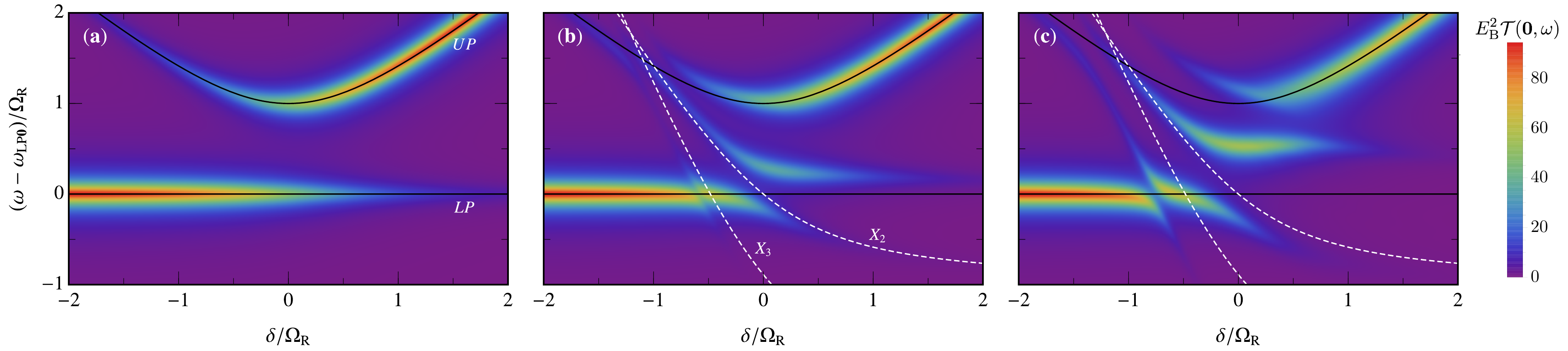}
\caption{Normal incidence pump-probe transmission ${\cal T} (\0,\omega)$ as a function of the photon-exciton detuning and the rescaled probe energy (relative to the LP energy) for increasing pump densities: (a) $n$=0, (b) $n=\mx\OR/16\pi$ and (c) $n=\mx\OR/4\pi$. In the experimentally realistic case of $\OR=3$meV, this corresponds to densities (b) $n=3\times10^{10}{\rm cm}^{-2}$ and (c) $n=1.25\times10^{11}{\rm cm}^{-2}$. In both cases we take $\eb=\OR$, and a broadening $\Gamma=\OR/10$.  The lower and upper polariton energies in the absence of the medium are shown as black solid lines. The dashed white lines indicate the locations of the vacuum biexciton ($X_2$) and triexciton ($X_3$) resonances at $\omega+\omega_{\text{LP}\0}= -\eb$ and $\omega+2\omega_{\text{LP}\0}= \varepsilon_T $, respectively, with triexciton energy $\varepsilon_T \simeq -2.4\eb$~\cite{supmat}.  }
\label{fig:densplots}
\end{figure*}

We model the $\up$-$\down$ interactions between excitons using a contact potential, which in momentum space is constant with strength $g$ up to a momentum cutoff $\Lambda$. This is reasonable, since typical polariton wavelengths $\sim1/\sqrt{m_{\text{C}} \OR}$ greatly exceed the exciton Bohr radius that sets the exciton-exciton interaction length scale~\cite{ciuti98:interaction,rochat00:prb}.  The exciton-polariton coupling in Eq.~\eqref{eq:hamil} is given by $g_{\k\k'} = g \cos\theta_\k \cos\theta_{\k'}$, with the Hopfield factor \cite{hopfield58}
\begin{align}
    \cos\theta_{\k}&=\frac{1}{\sqrt{2}}\sqrt{1+\frac{\omega_{\text{C}\k}-\omega_{\text{X}\k}}{\sqrt{(\omega_{\text{C}\k}-\omega_{\text{X}\k})^2+\OR^2}}} \, ,
\end{align} 
which corresponds to the exciton fraction in the LP state at a given momentum.  As is standard in two-dimensional quantum gases (see, e.g., Ref.~\cite{2Dreview}), the coupling constant and cutoff are related to the biexciton binding energy $\eb$ (which we define as positive) through the process of renormalization:
\begin{equation} 
    -\frac{1}{g 
    } = 
    \sum_\k^{k<\Lambda}\frac1{\eb+2\omega_{\rm{X}\k}} 
    =\frac{m_{\rm X}}{4\pi} \ln \left(\frac{\Lambda^2/\mx + \eb}{\eb}\right).
\label{eq:g}
\end{equation}
This treatment of the ultraviolet physics is justified as long as the biexciton size greatly exceeds that of the exciton, which is the case when the masses of the electron and hole making up the exciton are comparable~\cite{Ivanov1998a}.  Note that we neglect interactions in the medium for simplicity.  If they were included, the polariton dispersion would be modified to that of medium quasiparticles, accounting for normal and anomalous interactions in the medium. This would not qualitatively change the results below.

\paragraph*{Probe photon transmission.--} 
The transmission ${\cal T}(\k,\omega)$ of a photon at frequency $\omega$ and momentum $\k$ is related to the photon retarded Green's function~\cite{Ciuti2006} via ${\cal T}(\k,\omega)=|G_\text{C}(\k,\omega)|^2$, where we ignore a constant prefactor that only depends on the loss rate through the mirrors. To evaluate this, we note that only the exciton component of the impurity interacts with the medium. Then, in the exciton-photon basis, the impurity Green's function has the form of a matrix,
\begin{align}
\mathbf{G}(\k,\omega) & =
\begin{pmatrix}
G_\text{X}^{(0)}(\k,\omega)^{-1}-\Sigma_{\text{X}}(\k,\omega) & -\OR/2 \\
-\OR/2 & G_\text{C}^{(0)}(\k,\omega)^{-1}
\end{pmatrix}^{-1},
\label{eq:G}
\end{align}
where $G_\text{C}\equiv\mathbf{G}_{22}$. Here, the exciton and photon Green's functions in the absence of interactions are $G_\text{X,C}^{(0)}(\k,\omega)=1/( \omega-\omega_{\text{X,C}\k}+i0)$, respectively, where the frequency poles are shifted infinitesimally into the lower complex plane. Importantly, Eq.~\eqref{eq:G} is an exact relation within the Hamiltonian \eqref{eq:hamil}, which highlights how any approximation to the probe transmission arises from the calculation of the exciton self-energy $\Sigma_\text{X}$.

In the following, we evaluate the photon Green's function by using the truncated basis method (TBM)~\cite{Parish2016}. Within this approximation, the Hilbert space of impurity wave functions is restricted to describe only a finite number of excitations of the medium. The Green's function can be found (as discussed below) by summing over all eigenstates in this basis.  In the context of ultracold gases, such an approximation has been shown to successfully reproduce the experimentally observed spectral function of impurities immersed in a Bose-Einstein condensate~\cite{Jorgensen2016}, as well as the ground state~\cite{Chevy2006,Vlietinck2013} and coherent quantum dynamics of impurities in a Fermi sea~\cite{Cetina2016}. As such, the TBM is an appropriate approximation for the investigation of impurity physics, both in and out of equilibrium.

\paragraph*{Impurity wave function. --}
To capture the signatures of strong two- and three-point correlations in the probe transmission, we introduce a variational wave function containing terms where the 
impurity is dressed by up to two excitations of the medium:
\begin{multline}
  \ket{\Psi}=\left(\gamma_0 \hat{c}^\dag_{\0}+\alpha_0 \exciton^\dag_{\0} +\sum_\k\alpha_\k
    \exciton^\dag_{-\k}\hat{L}^\dag_{\k}
  \right.\\
   \left. 
    +\frac12\sum_{\k_1\k_2}\alpha_{\k_1\k_2}
    \exciton^\dag_{-\k_1-\k_2} \hat{L}^\dag_{\k_1} \hat{L}^\dag_{\k_2}
  \right)\ket{\Phi} .
\label{eq:wavefunc}
\end{multline}
Here $\ket{\Phi}$ is the coherent state describing the medium in the absence of the impurity, and we consider a $\sigma_-$ probe at normal incidence, where the total momentum imparted is zero. We take advantage of the fact that the large mass difference between photons and excitons acts to suppress terms in the wave function containing impurity photons at finite momentum --- i.e., terms such as $\gamma_\k \hat{c}^{\dag}_{-\k} \hat{L}^\dag_{\k}$ and $\gamma_{\k_1\k_2} \hat{c}^\dag_{-\k_1-\k_2} \hat{L}^\dag_{\k_1} \hat{L}^\dag_{\k_2}$ are far detuned in energy from the other terms in the wave function, and have thus been neglected. We then find the impurity spectrum by solving $\hat H\ket{\Psi}=E\ket{\Psi}$ within the truncated Hilbert space given by wave functions of the form~\eqref{eq:wavefunc}.  This procedure yields a set of coupled linear equations that we solve numerically~\cite{supmat}.

Within the TBM, once all eigenvalues and vectors of the linear equations are known, the photon Green's function can be written as~\cite{supmat}:
\begin{align}
G_{\rm C}(\0,\omega) & 
\simeq\sum_n\frac{|\gamma^{(n)}_0|^2}{\omega-E_n+i\Gamma}.
\label{eq:green}
\end{align}
The sum runs over all eigenstates within the truncated Hilbert space, picking out the weight of the photon term from each.  The factor $i\Gamma$ introduces broadening because of microcavity finite lifetime effects. For simplicity we take it to be independent of the state, which corresponds to considering equal exciton and photon lifetimes. This does not qualitatively affect the results of our work.

\paragraph*{Results.--} 
In Fig.~\ref{fig:densplots} we show our calculated normal incidence pump-probe transmission as a function of the photon-exciton detuning and the probe frequency.  In the limit of vanishing pump power, Fig.~\ref{fig:densplots}(a), the probe transmission is given by the single-particle LP and UP branches as expected~\cite{hopfield58,pekar58}, with the relative weights varying according to the photonic fraction of each branch. On increasing the pump strength, we observe first one and then two additional branches appearing with clear avoided crossings, as depicted in panels (b) and (c). This happens in the vicinity of where the LP and UP branches become resonant with either a biexciton ($X_2$) or a triexciton ($X_3$) state: Indeed, recalling that we set the $k=0$ exciton energy to zero, the crossings between solid and dashed lines correspond to the \textit{zero-density} resonance conditions $\omega^\ast+\omega_{{\rm LP}\0}=-\eb$ and $\omega^\ast+2\omega_{{\rm LP}\0}=\varepsilon_T$, where $\varepsilon_T$ is the vacuum triexciton energy, and $\omega^\ast \in \{\omega_{{\rm LP}\0}, \omega_{{\rm UP}\0}\}$.  The resonant behavior results in an intriguing transmission spectrum, where both lower and upper polaritons split into red-shifted attractive and blue-shifted repulsive polaronic quasiparticle branches due to the $X_3$ and $X_2$ resonances.  Furthermore, at sufficiently large densities, we see that the two LP repulsive branches smoothly evolve into the corresponding attractive and repulsive branches of the UP state.

It is important to distinguish the nature of the polaron state we describe here from the mean-field coupled-channel picture described elsewhere~\cite{Takemura2014,Takemura2017}, which, at low densities, produces a qualitatively similar spectrum.  In the coupled-channel model, there is an anticrossing between the polariton branches and a pre-formed molecular state.  By contrast, the $X_2$ splitting described in this Letter is a beyond-mean-field many-body effect due to two-point correlations which are enhanced by the biexciton resonance.  Similarly, the appearance of additional branches at higher densities demonstrates the emergence of many-body three-point correlated states. Indeed, we see that the $X_3$ resonance position gets rapidly shifted from the vacuum triexciton energy when increasing the density.  Note that our model is likely to overestimate the magnitude of the triexciton energy $|\varepsilon_T|$, since we have neglected the repulsion between $\up$ excitons. However, we can show that the triexciton remains bound even when there is an effective three-body repulsion (which mimics the $\up$-$\up$ repulsion~\cite{yoshida2018}), and the triexciton binding energy only weakly depends on this repulsion~\cite{supmat}.

\begin{figure}[t] \centering \includegraphics[width=.9\columnwidth]{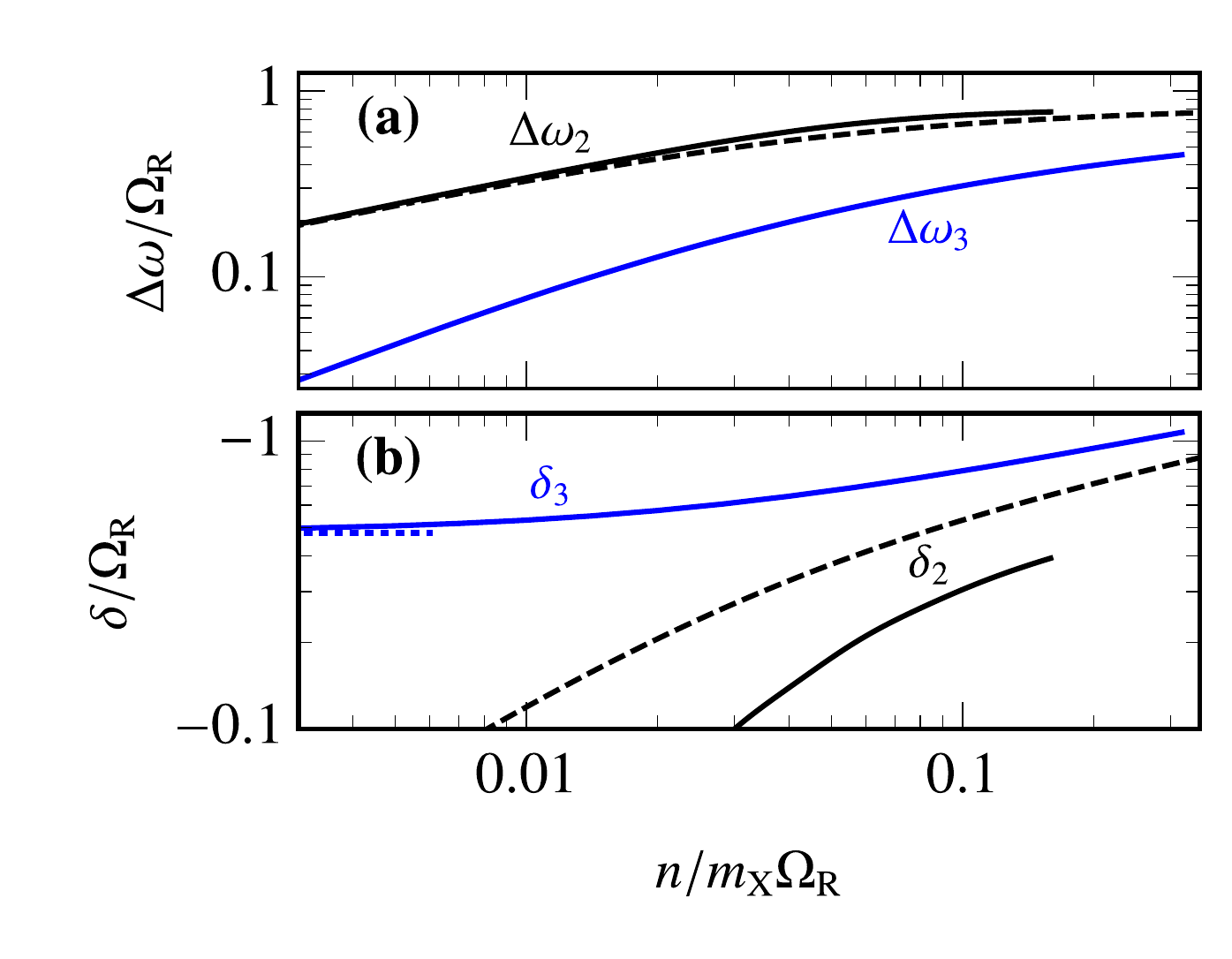} 
\caption{(a) Minimal splitting between LP quasiparticle branches in the transmission spectrum, and (b) photon-exciton detuning at the minimal splitting. The splitting $\Delta \omega_3$ and detuning $\delta_3$ for the lowest two branches --- originating from the triexciton resonance --- are shown as solid blue lines, where $\delta_3\to -0.48 \OR$ (dotted line) in the limit $n\to0$ due to the triexciton state.  The splitting $\Delta \omega_2$ and detuning $\delta_2$ for the biexciton resonance are shown as solid black lines.  The dashed black lines depict the corresponding results calculated when including only two-point correlations (i.e., the Hilbert space with at most one excitation of the medium).}
\label{fig:split}
\end{figure}

In order to quantify the density dependence of the two $X_2$ and $X_3$ resonances for the lower polariton, we evaluate in Fig.~\ref{fig:split} the minimal splittings $\Delta \omega_{2,3}$ between repulsive and attractive branches and the corresponding detunings $\delta_{2,3}$ at which these anticrossings occur when the finite lifetime broadening $\Gamma$ can be neglected~\cite{supmat}.  In the low-density limit, one can formally show that the minimal splitting due to the $X_2$ resonance has the form $\Delta \omega_2 \sim \cos\theta_0\sqrt{n\eb/\mx}$~\cite{supmat}. This behavior is captured using two-point correlations only, and indeed we see in Fig.~\ref{fig:split}(a) that two-point correlations dominate even at higher densities. However, the shift in the detuning $\delta_2$ is a higher order density effect which can be affected by three-point correlations, as illustrated in Fig.~\ref{fig:split}(b). For the $X_3$ resonance, the splitting $\Delta \omega_3$ shown in Fig.~\ref{fig:split}(a) approaches a linear scaling with $n$ as $n \to 0$. In this case, one can show that the energy shift of the attractive branch scales linearly with $n$ at low densities, while the repulsive branch only shifts upwards once $\delta_3$ moves away from the vacuum resonance position~\cite{supmat}. Note that, in the presence of broadening, a given splitting $\Delta \omega$ is only visible when $\Delta \omega\gtrsim \Gamma$.

\paragraph*{Implications for experiments.--}
As previously mentioned, the pump-probe protocol employed in the experiments by Takemura {\em et al.}~\cite{Takemura2014,Takemura2017} is similar to our impurity scenario. However, Ref.~\cite{Takemura2017} focused on a low density of $\up$ polaritons, where no splitting can be observed, while the experiment of Ref.~\cite{Takemura2014} employed a broad pump that populated \textit{both} $\up$ LP and UP branches.  Nevertheless, if we take $\OR=\eb=3$meV, then the parameters chosen for Fig.~\ref{fig:densplots}(c) correspond to a density of $n=1.25\times10^{11}\text{cm}^{-2}$, which approximately matches the parameters of Fig.~3 in Ref.~\cite{Takemura2014}. Here, the splitting of the lower polariton close to the biexciton resonance was analyzed~\cite{Takemura2014}. Qualitatively, our results for the attractive and repulsive energy shifts agree; however the measured energy shifts are somewhat smaller than what we find.  This is likely to be due to the broad range of $\up$ states populated in Ref.~\cite{Takemura2014}, which will tend to wash out the effect of the resonances compared to when the bosonic medium is a macroscopically occupied single-particle state.

\paragraph*{Conclusions and outlook.--} 
In this Letter, we have shown how many-body correlations in the exciton-polariton system can be directly accessed using pump-probe spectroscopy. Such measurements should be simpler than the sophisticated multi-dimensional optical spectroscopy techniques employed in, e.g., Ref.~\cite{Wen2013b}, which require multiple phase-stable optical pulses with controllable delays.  Furthermore, depending on the material parameters, there is even the possibility of overlapping biexciton and triexciton resonances, where both two- and three-point correlations are enhanced~\cite{supmat}.  Direct probes of many-body correlated states can also provide stringent bounds on the nature and spin structure of the polariton-polariton interaction; they can thus provide a complementary approach to resolve questions over the size of this interaction~\cite{ciuti98:interaction,rochat00:prb,Sun2017}.

\acknowledgements We are grateful to  A.~\.{I}mamo\u{g}lu for useful discussions.  JL and MMP acknowledge support from the Australian Research Council Centre of Excellence in Future Low-Energy Electronics Technologies (CE170100039).  JL is also supported through the Australian Research Council Future Fellowship FT160100244.  FMM acknowledges financial support from the Ministerio de Econom\'ia y Competitividad (MINECO), projects No.~MAT2014-53119-C2-1-R and No.~MAT2017-83772-R.  JK acknowledges financial support from EPSRC program ``Hybrid Polaritonics'' (EP/M025330/1).

\bibliography{cavity_TMD_refs}


\renewcommand{\theequation}{S\arabic{equation}}
\renewcommand{\thefigure}{S\arabic{figure}}

\onecolumngrid

\newpage

\setcounter{equation}{0}
\setcounter{figure}{0}
\setcounter{page}{1}

\section*{SUPPLEMENTAL MATERIAL: ``Spectroscopic probes of quantum many-body correlations in polariton microcavities''}

\begin{center}
Jesper Levinsen$^{1,2}$, Francesca Maria Marchetti$^{3}$, Jonathan Keeling$^{4}$, and Meera M.~Parish$^{1,2}$,\\
\emph{\small $^1$School of Physics and Astronomy, Monash University, Victoria 3800, Australia}\\
\emph{\small $^2$ARC Centre of Excellence in Future Low-Energy Electronics Technologies, Monash University, Victoria 3800, Australia}\\
\emph{\small $^3$Departamento de F\'isica Te\'orica de la Materia Condensada \& Condensed Matter Physics Center (IFIMAC), Universidad Aut\'onoma de Madrid, Madrid 28049, Spain}\\
\emph{\small $^4$SUPA, School of Physics and Astronomy, University of St Andrews, St Andrews, KY16 9SS, United Kingdom}
\end{center}

\section{Hamiltonian}
To model a polariton impurity with spin $\down$ (generated by a weak $\sigma_-$ probe) immersed in the medium of spin $\up$ polaritons (generated by a $\sigma_+$ pump), we start from a Hamiltonian in the exciton ($\exciton_{\k\sigma}^{}$) and cavity photon ($\hat{c}_{\k\sigma}^{}$) basis (as in the main text, $\hbar$ and system area are set to $1$):
\begin{equation}
  \hat{H}_{\rm ex-ph} = \sum_{\k, \sigma = \uparrow,\downarrow}
  \omega_{\text{X}\k} 
  \exciton_{\k\sigma}^\dag
  \exciton_{\k\sigma}^{} + g\sum_{\k,
    \k', \vect{q}} 
\exciton_{\k \uparrow}^\dag
\exciton_{\q-\k \downarrow}^\dag \exciton_{\q-\k'
  \downarrow}^{}
\exciton_{\k' \uparrow}^{}
  + \sum_{\k, \sigma = \uparrow,\downarrow}
  \omega_{\text{C}\k} 
  \hat{c}_{\k\sigma}^\dag
  \hat{c}_{\k\sigma}^{} + \frac{\OR}{2} \sum_{\k, \sigma =
    \uparrow,\downarrow} \left(\exciton_{\k\sigma}^\dag
    \hat{c}_{\k\sigma}^{} + \text{h.c.}\right) \; .
\label{eq:hamsm}
\end{equation}
Here, we have neglected the interaction in the medium (spin $\up$ excitons) and approximated the interaction between excitons of opposite spin as a contact interaction with strength $g$ renormalized via the biexciton binding energy $\eb$ according to Eq.~\eqref{eq:g}.  Approximating the exciton-exciton interaction as contact is justified by the typical polariton wavelengths $\ell = 1/\sqrt{m_{\text{C}} \OR}$ being much larger than the exciton Bohr radius $a_{0}$.

As explained in the main text, the $\sigma_+$ pump resonantly injects spin $\up$ polaritons at normal incidence, with energy equal to $\omega_{\text{LP}\0}$. For this reason, it is profitable to rotate the exciton and photon $\up$ states into the lower (LP) and upper polariton (UP) basis,
\begin{align}
\begin{pmatrix} 
\hat{L}_{\k\up} \\ \hat{U}_{\k\up}
\end{pmatrix}
=
\begin{pmatrix}
  \cos\theta_{\k} & \sin\theta_{\k}\\-\sin\theta_{\k} &
  \cos\theta_{\k}
\end{pmatrix}
\begin{pmatrix}
  \exciton_{\k\up}^{} \\ \hat{c}_{\k\up}^{}
\end{pmatrix} \; ,
\end{align}
where the Hopfield factors are given by:
\begin{align}
\cos\theta_{\k} & = \frac1{\sqrt{2}}
\sqrt{1 + \frac{\omega_{\text{C}\k}-\omega_{\text{X}\k}}{\sqrt{(\omega_{\text{C}\k}-\omega_{\text{X}\k})^2+\OR^2}}} & 
\sin\theta_{\k} &= \frac1{\sqrt{2}}
\sqrt{1 - \frac{\omega_{\text{C}\k}-\omega_{\text{X}\k}}{\sqrt{(\omega_{\text{C}\k}-\omega_{\text{X}\k})^2+\OR^2}}} \; .
\end{align}
We can then rewrite the Hamiltonian~\eqref{eq:hamsm} in terms of the LP polariton operators $\hat{L}_{\k\up}$ by neglecting the contribution of the UP states $\hat{U}_{\k\up}$ for the spin $\up$ particles, which are barely occupied by the pump. In the same spirit, we explicitly separate the contribution of the macroscopically occupied $\k=\0$ state from the $\k\ne\0$ excitations by substituting $\hat{L}_{\k\up} \mapsto \sqrt{n}\delta_{\k=\0}+\hat{L}_{\k\ne\0\up}$, where $n$ is the medium density. By measuring energies with respect to the energy of the medium, $\omega_{\text{LP}\0} n$, in the absence of the impurity and excitations, we then obtain the expression~\eqref{eq:hamil} in the main text (note that in the main text, for brevity, we have suppressed the spin-dependence of the operators, i.e., $\exciton_{\k\down}^{} \mapsto \exciton_{\k}^{}$, $\hat{c}_{\k\down}^{} \mapsto \hat{c}_{\k}^{}$, and $\hat{L}_{\k\up}^{} \mapsto \hat{L}_{\k}^{}$).

\section{Polaron wave function and truncated basis method}

For a given momentum $\Q$, the dressed $\down$ impurity state can be described by the following variational wave function,
\begin{multline}
  \ket{\Psi_\Q}=\left(\alpha_{\Q;0} \exciton^\dag_{\Q\down} +\sum_\k\alpha_{\Q;\k}
    \exciton^\dag_{\Q-\k\down} \hat{L}^\dag_{\k\up}
    +\frac12\sum_{\k_1\k_2}\alpha_{\Q;\k_1\k_2}
    \exciton^\dag_{\Q-\k_1-\k_2,\down} \hat{L}^\dag_{\k_1\up} \hat{L}^\dag_{\k_2\up} \right.\\
  + \left.  \gamma_{\Q;0} \hat{c}^\dag_{\Q\down} +\sum_\k \gamma_{\Q;\k}
    \hat{c}^\dag_{\Q-\k\down} \hat{L}^\dag_{\k\up}
    +\frac12\sum_{\k_1\k_2}  \gamma_{\Q;\k_1\k_2}
    \hat{c}^\dag_{\Q-\k_1-\k_2,\down} \hat{L}^\dag_{\k_1\up} \hat{L}^\dag_{\k_2\up}
  \right)\ket{\Phi}.
\label{eq:fulwf}
\end{multline}
where $\ket{\Phi}$ is the state of the medium following the resonant pumping at zero momentum, i.e., it is the state that satisfies $\hat{L}_{\k\up}\ket{\Phi} = 0$. This initial state of course coincides with the coherent state of a lower polariton Bose-Einstein condensate. Note that we require $\alpha_{\Q;\k\k'} = \alpha_{\Q;\k'\k}$ and $\gamma_{\Q;\k\k'} = \gamma_{\Q;\k'\k}$ in order to satisfy Bose statistics.  The first (second) line of Eq.~\eqref{eq:fulwf} describes the exciton (photon) component of the bare $\down$ impurity and its dressing by both one and two medium excitations.  We thus include two-point correlations (via the $\alpha_{\Q;\k}$ and $\gamma_{\Q;\k}$ terms) as well as three-point correlations (via the $\alpha_{\Q;\k_1\k_2}$ and $\gamma_{\Q;\k_1 \k_2}$ terms) between the $\down$-impurity and the reservoir of $\up$-polaritons.  Finally, the normalization condition requires that, for each value of the momentum $\Q$,
\begin{equation}
    1 = \langle \Psi_\Q \ket{\Psi_\Q} = |\alpha_{\Q;0}|^2 + \sum_{\k} |\alpha_{\Q;\k}|^2 + \frac12 \sum_{\k_1 \k_2} |\alpha_{\Q;\k_1\k_2}|^2 + | \gamma_{\Q;0}|^2 + \sum_{\k} |\gamma_{\Q;\k}|^2 + \frac12 \sum_{\k_1 \k_2} | \gamma_{\Q;\k_1\k_2}|^2\; .
\end{equation}

\subsection{Probing at normal incidence, $\Q=\0$}
Let us first consider a photon probe at normal incidence, $\Q=\0$. We then minimize the equation $\bra{\Psi_\0} (\hat H - E) \ket{\Psi_\0}$ with respect to the variational parameters $\left\{\alpha_0, \gamma_0, \alpha_\k,
  \gamma_\k,
  \alpha_{\k_1\k_2},
  \gamma_{\k_1\k_2}
\right\}$ (for simplicity, we drop the $\Q=\0$ momentum subscripts). This procedure yields the set of linear equations
\begin{subequations} \label{eq:varpara}
\begin{align}
    E \alpha_0  = & \frac{\OR}{2} \gamma_0 + g \sqrt{n} \cos\theta_\0 \sum_\q \cos\theta_\q \alpha_\q \\
    E \gamma_0  = & \delta \gamma_0 + \frac{\OR}2 \alpha_0 \\ 
    E \alpha_\k  = & (\omega_{\text{X}\k}+ \omega_{\text{LP}\k}-\omega_{\text{LP}\0})\alpha_\k + \frac{\OR}{2} \gamma_\k + g \sqrt{n} \cos\theta_\0 \cos\theta_\k \alpha_0 + g \cos\theta_\k \sum_{\k'} \cos\theta_{\k'} \alpha_{\k'}  + g \sqrt{n} \cos\theta_\0 \sum_{\k'} \cos\theta_{\k'} \alpha_{\k\k'}\\
    E \gamma_\k  =  & (\omega_{\text{C}\k}+ \omega_{\text{LP}\k}-\omega_{\text{LP}\0} )\gamma_\k + \frac{\OR}{2} \alpha_\k \\ \notag
    E \alpha_{\k_1\k_2} = & (\omega_{\text{X}\k_1 + \k_2}+ \omega_{\text{LP}\k_1} + \omega_{\text{LP}\k_2}-2\omega_{\text{LP}\0} ) \alpha_{\k_1\k_2} + \frac{\OR}{2} \gamma_{\k_1\k_2} + g \sqrt{n} \cos\theta_\0 \left(\cos\theta_{\k_1} \alpha_{\k_2} + \cos\theta_{\k_2} \alpha_{\k_1} \right)\\
    & + g \cos\theta_{\k_2} \sum_{\k'} \cos\theta_{\k'} \alpha_{\k_1 \k'} + g \cos\theta_{\k_1} \sum_{\k'} \cos\theta_{\k'} \alpha_{\k' \k_2}
    \\ 
    E \gamma_{\k_1\k_2}  = & (\omega_{\text{C}\k_1 + \k_2}+ \omega_{\text{LP}\k_1} + \omega_{\text{LP}\k_2} -2\omega_{\text{LP}\0} ) \gamma_{\k_1\k_2} + \frac{\OR}{2} \alpha_{\k_1\k_2} \; .
\end{align}
\end{subequations}
We numerically solve these coupled equations by considering them as an eigenvalue problem on a discrete 2D grid in momentum $\k=(k,\varphi_k)$ space, and we have carefully checked the convergence of our results with respect to the number of grid points.

The photonic and excitonic components of the impurity Green's functions can be written as
\begin{equation}
    G_{\text{C},\text{X}} (\0,\omega) = \int d\omega' \frac{Z_{\text{C},\text{X}} (\omega')}{\omega - \omega' + i0}\; .
\end{equation}
Here, $Z_{\text{C},\text{X}} (\omega)$ is the corresponding quasiparticle residue at frequency $\omega$, i.e., it is the overlap with the non-interacting state. Within our truncated basis method, we replace the integral over all frequencies by a sum over all the eigenstates of Eq.~\eqref{eq:varpara}~\cite{Parish2016}. We then have
\begin{align}
    G_{\text{C}} (\0,\omega) &\simeq \sum_n \frac{\left|\gamma_0^{(n)}\right|^2}{\omega-E_n+i0}
    &
    G_{\text{X}} (\0,\omega) &\simeq \sum_n \frac{\left|\alpha_0^{(n)}\right|^2}{\omega-E_n+i0}\; ,
\end{align}
where $E_n$ is the eigenvalue of the $n$'th eigenstate of Eq.~\eqref{eq:varpara}, and $\alpha_0^{(n)}$ and $\gamma_0^{(n)}$ are the values of $\alpha_0$ and $\gamma_0$ in that state. This procedure yields a series of discrete peaks, and in order to obtain a continuous transmission spectrum we replace $i0\to i\Gamma$, which models a finite lifetime in the microcavity. Evaluating $\mathcal{T} (\0,\omega) = |G_{\text{C}} (\0,\omega)|^2$ in this manner, we arrive at Fig.~\ref{fig:densplots} of the main text.

Because of the large mass difference between photons and excitons, $\mc/\mx\ll1$ (throughout the manuscript we choose $\mc = 10^{-4}\mx$), in the variational Ansatz~\eqref{eq:fulwf} we can neglect the contribution of the excited modes in the photonic component, $\gamma_\k$ and $\gamma_{\k_1\k_2}$, obtaining the wave function of the main text~\eqref{eq:wavefunc} and a reduced set of equations to solve:
\begin{subequations} \label{eq:varpar2}
\begin{align}
    E \alpha_0  = & \frac{\OR}{2} \gamma_0 + g \sqrt{n} \cos\theta_\0 \sum_\q \cos\theta_\q \alpha_\q \\
    E \gamma_0  = & \delta \gamma_0 + \frac{\OR}2 \alpha_0 \\ 
    E \alpha_\k  = & (\omega_{\text{X}\k}+ \omega_{\text{LP}\k}-\omega_{\text{LP}\0})\alpha_\k + g \sqrt{n} \cos\theta_\0 \cos\theta_\k \alpha_0 + g \cos\theta_\k \sum_{\k'} \cos\theta_{\k'} \alpha_{\k'}  + g \sqrt{n} \cos\theta_\0 \sum_{\k'} \cos\theta_{\k'} \alpha_{\k\k'}\\
    E \alpha_{\k_1\k_2} = & (\omega_{\text{X}\k_1 + \k_2}+ \omega_{\text{LP}\k_1} + \omega_{\text{LP}\k_2}-2\omega_{\text{LP}\0} ) \alpha_{\k_1\k_2} + g \sqrt{n} \cos\theta_\0 \left(\cos\theta_{\k_1} \alpha_{\k_2} + \cos\theta_{\k_2} \alpha_{\k_1} \right)\\
    & + g \cos\theta_{\k_2} \sum_{\k'} \cos\theta_{\k'} \alpha_{\k_1 \k'} + g \cos\theta_{\k_1} \sum_{\k'} \cos\theta_{\k'} \alpha_{\k' \k_2} \; .
\end{align}
\end{subequations}
Further, in the same limit $\mc/\mx\ll1$, we can expand around the limit of vanishing photon mass. This means that for non-vanishing momentum $\k$ we can take $\omega_{{\rm LP}\k}\simeq\omega_{{\rm X}\k}$ and $\cos\theta_\k\simeq1$. We have checked that both approximations makes no visible quantitative difference to the results displayed in Figs.~\ref{fig:densplots} and~\ref{fig:split} in the main text.


\subsection{Splitting between quasiparticle branches}

\begin{figure}[h] \centering \includegraphics[width=.5\columnwidth]{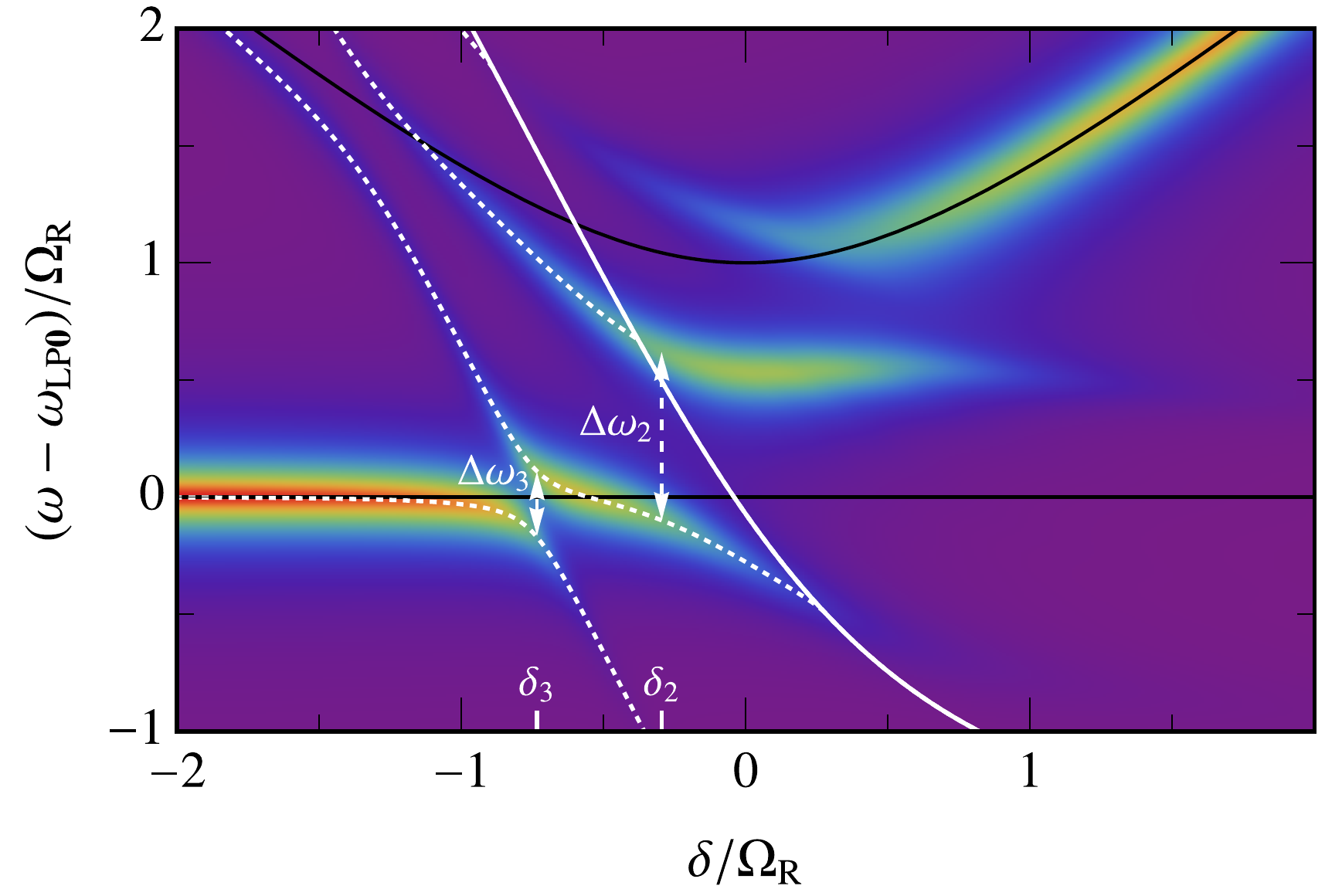} 
\caption{Illustration of our calculation of the locations, $\delta_2$ and $\delta_3$ (marked on the $x$-axis), and magnitudes, $\Delta \omega_2$ and $\Delta \omega_3$ (denoted by arrows) of the minimal splitting between quasiparticle branches. The system parameters are fixed as in Fig.~\ref{fig:densplots}(c) of the main text. The four lowest lying eigenvalues are marked as white dotted lines, while the edge of the continuum is above the white solid line.}
\label{fig:extract}
\end{figure}

In Fig.~\ref{fig:extract} we illustrate how we extract the minimal splitting between the quasiparticle branches at a given pump density $n$, yielding the results shown in Fig.~\ref{fig:split} of the main text. First, we evaluate the locations of the three lowest lying maxima in the spectrum at a fixed detuning $\delta$, and then we find the location and magnitude of the minimum splitting between two neighboring maxima. Within the truncated Hilbert space, the spectrum is discrete for frequencies below the continuum, as illustrated in Fig.~\ref{fig:extract}.  Therefore, the location of the lowest lying transmission maxima (in practice, the two lowest lying maxima) can be easily extracted from the energy eigenvalues of Eq.~\eqref{eq:varpar2} in the limit $\Gamma\to0$. The third maximum is in the continuum, and we evaluate its location by taking $\Gamma=0.1\OR$. Thus, in Fig.~\ref{fig:split} of the main text, the black solid line is (weakly) dependent on the value of $\Gamma$ chosen, whereas the blue solid and black dashed lines both correspond to the limit $\Gamma\to0$.

\subsection{Renormalized equations and the low-density limit}

We can rewrite our variational equations in an explicitly cutoff independent fashion by defining 
\begin{subequations} \label{eq:eta}
\begin{align}
    \eta & = g \sum_\k \cos\theta_\k \alpha_\k \\
    \xi_\k & = g \sum_{\k'} \cos\theta_{\k'} \alpha_{\k\k'},
\end{align}
\end{subequations}
where we focus on the $\Q=\0$ case.  Eliminating the $\gamma_0$, $\gamma_\k$ and $\gamma_{\k_1\k_2}$ terms from Eqs.~\eqref{eq:varpara} and then taking the limit $\Lambda \to \infty$, we obtain
\begin{align} \label{eq:1st}
\left(\frac{1}g + \sum_\k \frac{\cos^2\theta_\k}{-E + E_\k} \right) \eta = & \frac{n  \cos^2\theta_\0}{E + \frac{\Omega_{\rm R}^2}{4}\frac{1}{\delta-E}} \eta 
- \sqrt{n} \cos\theta_\0 \sum_\k \frac{\cos\theta_\k \, \xi_\k}{-E + E_\k}, \\ \label{eq:2nd}
\left(\frac{1}g + \sum_{\k'} \frac{\cos^2\theta_{\k'}}{-E + E_{\k\k'}} +\frac{n\cos^2\theta_\0}{-E+E_\k}\right) \xi_\k = &  \, 
\frac{\sqrt{n}\cos\theta_\0\cos\theta_\k } 
{E-E_\k}\eta + \cos\theta_\k \sum_{\k'} \frac{\cos\theta_{\k'}\xi_{\k'}}{E -E_{\k\k'}}, 
\end{align}
where we have used the fact that the quantities in Eq.~\eqref{eq:eta} are finite in this limit, and we have defined
\begin{align}
    E_\k = &  \omega_{\text{X}\vect{k}}+ \omega_{\text{LP}\vect{k}} -\omega_{{\rm LP}\0} - \frac{\Omega_{\rm R}^2}{4} \frac{1}{-E + \omega_{\text{C}\vect{k}}+ \omega_{\text{LP}\vect{k}}-\omega_{{\rm LP}\0} }, \\
    E_{\k_1 \k_2} = & \omega_{\text{X}\vect{k}_1 + \k_2}+ \omega_{\text{LP}\vect{k}_1} + \omega_{\text{LP}\vect{k}_2} -2\omega_{{\rm LP}\0} - \frac{\Omega_{\rm R}^2}{4} \frac{1}{-E + \omega_{\text{C}\vect{k}_1+\k_2}+ \omega_{\text{LP}\vect{k}_1} + \omega_{\text{LP}\vect{k}_2}-2\omega_{{\rm LP}\0} }.
\end{align}
The solution to these equations directly gives the energy of the attractive impurity branch (the impurity ``ground state'') that lies below the lower polariton state, as well as the energy of all states that are below the continuum. In principle, all excited states are also encoded in these equations; however due to the complicated pole and branch cut structure for energies in the continuum, these are not easy to extract. We therefore extract the spectrum using the linear equations in Eq.~\eqref{eq:varpar2}.  In the limit of vanishing density, $n \to 0$, and vanishing photon mass, Eqs.~\eqref{eq:1st} and \eqref{eq:2nd} give the equations for biexciton and triexciton bound states, i.e., Eq.~\eqref{eq:g} in the main text and Eq.~\eqref{eq:2dtrimer} below, respectively.

For small but finite densities, away from the triexciton resonance, we can neglect the last term in Eq.~\eqref{eq:1st} and obtain an implicit equation for the quasiparticle energies when $E+\omega_{\text{LP}\0} < 0$:
\begin{align} \label{eq:low}
    E \simeq \frac{\OR^2}{4} \frac{1}{E-\delta} + \frac{n}{\mx} \cos^2\theta_\0  \frac{4\pi}{\ln\left(\frac{\eb}{-E-\omega_{\text{LP}\0}}\right)} , 
\end{align}
where we have used the fact that $\mc/\mx \ll 1$.  If one instead considers a mean-field two-channel approach, where the biexciton is treated as a structureless particle like in Refs.~\cite{Takemura2014,Takemura2017}, one obtains the implicit equation
\begin{align} \label{eq:Takemura}
    E \simeq \frac{\OR^2}{4} \frac{1}{E-\delta} + n 
    \cos^2\theta_\0  \frac{
    g^2_{\rm{BX}}}{E+\omega_{\text{LP}\0}+\eb} , 
\end{align}
where $g_{\text{B}\text{X}}$ is the effective coupling to the biexciton state. This amounts to approximating the exciton T-matrix $T_{\text{X}\text{X}}(E) = \frac{4\pi}{\mx} \frac{1}{\ln\left(\frac{\eb}{-E-\omega_{\text{LP}\0}}\right)} \simeq \frac{g^2_{\rm BX}}{E+\omega_{\text{LP}\0}+\eb}$ which is only accurate at the vacuum biexciton resonance, with $g^2_{\rm BX} = 4\pi\eb/\mx$. As a result, there are fundamental differences between their behavior in general: Eq.~\eqref{eq:Takemura} always has three distinct solutions, while Eq.~\eqref{eq:low} features a branch cut corresponding to a continuum of unbound states.  As a consequence, within these models, even the LP quasiparticle branches behave differently at high density $n$.  Note that, because of the low photon-exciton mass ratio, $m_{\text{C}}=10^{-4} m_{\text{X}}$, the polariton T-matrix is to a very good approximation the exciton T-matrix.

At the vacuum biexciton resonance of the lower polariton, where $2\omega_{\text{LP}\0} \approx -\eb$, a low-density expansion of Eq.~\eqref{eq:low} yields the energies of the attractive ($-$) and repulsive ($+$) branches at leading order in the density:
\begin{align}\label{eq:low2}
 E_\pm \simeq \omega_{\text{LP}\0} \pm 2 \cos\theta_0 \sqrt{\frac{\pi n \eb}{\mx} \left(\frac{\delta - \omega_{\text{LP}\0}}{\delta -2\omega_{\text{LP}\0}} \right)}.  
\end{align}
This yields the splitting $\Delta \omega_2 = E_+ - E_- \sim \cos\theta_0\sqrt{n\eb/\mx}$ at low densities, as quoted in the main text.

For the triexciton resonance of the lower polariton, we must consider both two- and three-point correlations, as encapsulated in Eqs.~\eqref{eq:1st} and \eqref{eq:2nd}. In this case, for the vacuum resonance condition $3\omega_{\text{LP}\0} = \varepsilon_T$ (with $\varepsilon_T$ the triexciton energy), we only obtain an energy shift for the attractive branch
\begin{align} \label{eq:low3}
    E - \omega_{\text{LP}\0} \sim - \frac{n\cos^2\theta_0}{\mx} \frac{1}{\ln\left(\frac{2|\varepsilon_T|}{3\eb}\right)}. 
\end{align}
To obtain an energy shift of the repulsive branch, we already need to consider detunings away from the vacuum triexciton resonance, which is consistent with what we see in Fig.~\ref{fig:densplots}. This complicates the behavior of the splitting $\Delta \omega_3$ at the triexciton resonance as the density is increased.  However, we can see from Eq.~\eqref{eq:low3} that the magnitude of the energy shift (and associated splitting) increases as we approach the condition $2|\varepsilon_T|= 3 \eb$, which corresponds to overlapping biexciton and triexciton resonances in the zero-density limit.

\subsection{Density dependence of transmission spectrum}
%
\begin{figure}[h] \centering \includegraphics[width=.9\columnwidth]{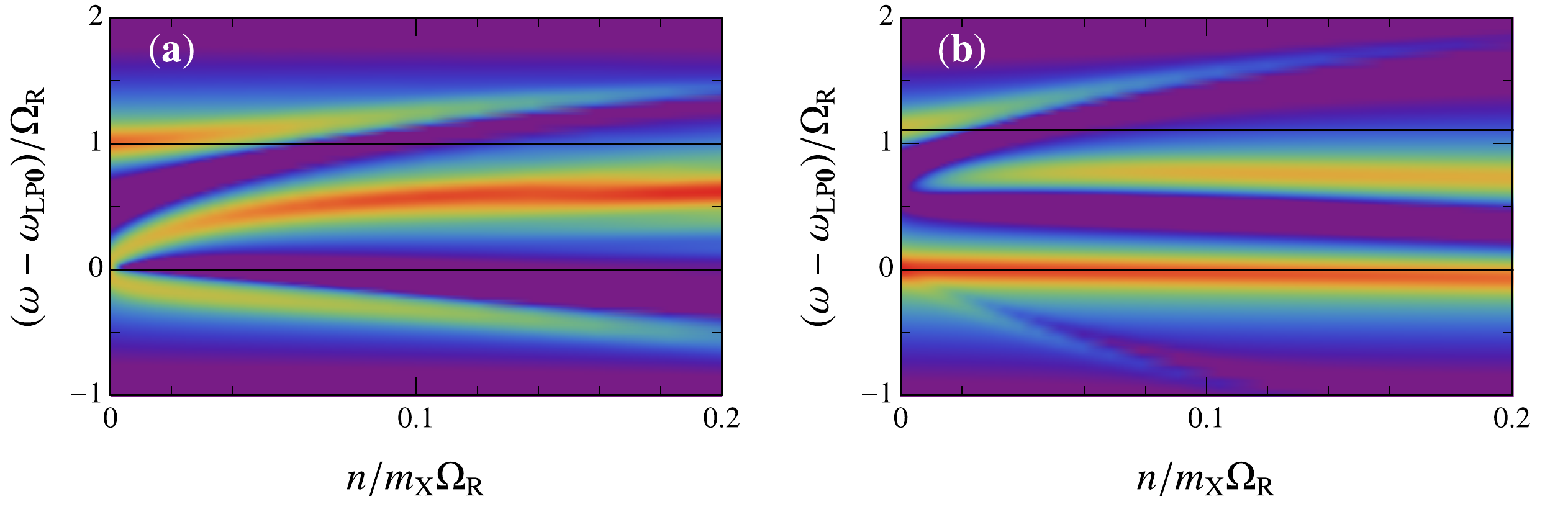} 
\caption{Density dependence of the normal probe transmission spectrum at fixed photon-exciton detuning. We show our results for (a) detuning $\delta=0$ corresponding to the crossing of the vacuum biexciton state with the lower polariton, and (b) detuning $\delta=-0.48\OR$ corresponding to the crossing of the vacuum triexciton state with the lower polariton. These crossings are illustrated in Fig.~2(b,c) of the main text.  We take $\eb=\OR$ as in the main text and a linewidth $\Gamma = \OR/20$. To clearly show the branches, we have used a log scale and we only show the transmission above a threshold. }
\label{fig:density}
\end{figure}

In Fig.~\ref{fig:density}, we plot the density dependence of the normal probe transmission spectrum at two fixed photon-exciton detunings. In both panels, we can observe the repulsion between branches as the density increases.  In particular, in panel (a) we consider a detuning $\delta=0$ corresponding to the crossing of the vacuum biexciton state with the lower polariton. Here, we observe that, at low density, the second and third lowest branches both originate from the LP branch, and their separation (splitting) increases approximately as $\sqrt{n}$, as expected from Eq.~\eqref{eq:low2}. At this detuning there is also a lower branch arising from three-point correlations; however it is below the frequency range plotted and has negligible spectral weight.  In panel (b) we instead take $\delta=-0.48\OR$, which corresponds to the crossing of the vacuum triexciton state with the lower polariton branch. Here the second lowest branch remains close to the LP energy, while the energy of the lowest (attractive) branch is shifted in a manner that scales linearly with $n$ at low density, in agreement with Eq.~\eqref{eq:low3}.

\subsection{Probing at a finite angle, $\Q\neq \0$}
If the $\sigma_-$ probe is incident at a finite angle (and hence a finite momentum $\Q$), we evaluate the probe spectrum of Fig.~\ref{fig:dispersion} in the low density regime where three-point correlations can be neglected. Thus, we consider states of the form
\begin{equation}
  \ket{\Psi_\Q} \simeq\left(\alpha_{\Q;0} \exciton^\dag_{\Q\down} +\sum_\k\alpha_{\Q;\k}
    \exciton^\dag_{\Q-\k\down} \hat{L}^\dag_{\k\up} + \gamma_{\Q;0} \hat{c}^\dag_{\Q\down} +\sum_\k \gamma_{\Q;\k}
    \hat{c}^\dag_{\Q-\k\down} \hat{L}^\dag_{\k\up}
  \right)\ket{\Phi}\; .
\label{eq:appwf}
\end{equation}
Now, the equations to solve become:
\begin{subequations} \label{eq:finiteQ}
\begin{align}
    E \alpha_{\Q;0}  = & \omega_{\text{X}\Q} \alpha_{\Q;0} +\frac{\OR}{2}  \gamma_{\Q;0} + g \sqrt{n} \cos\theta_\0 \sum_\q \cos\theta_\q \alpha_{\Q;\q} \\
    E  \gamma_{\Q;0}  = & \omega_{\text{C}\Q}  \gamma_{\Q;0} + \frac{\OR}2 \alpha_{\Q;0} \\ 
    E \alpha_{\Q;\k}  = & (\omega_{\text{X}\k-\Q}+ \omega_{\text{LP}\k}-\omega_{\text{LP}\0})\alpha_{\Q;\k} + \frac{\OR}{2} \gamma_{\Q;\k} + g \sqrt{n} \cos\theta_\0 \cos\theta_\k \alpha_{\Q;0} + g \cos\theta_\k \sum_{\k'} \cos\theta_{\k'} \alpha_{\Q;\k'} \\
    E \gamma_{\Q;\k}  =  & (\omega_{\text{C}\k-\Q}+ \omega_{\text{LP}\k}-\omega_{\text{LP}\0} )\gamma_{\Q;\k} + \frac{\OR}{2} \alpha_{\Q;\k} \; .
\end{align}
\end{subequations}
Consequently, we arrive at the photon Green's function
\begin{align}
    G_{\text{C}} (\k,\omega) \simeq \sum_n \frac{\left|\gamma_{\k;0}^{(n)}\right|^2}{\omega-E_n+i0},
\end{align}
from which we determine ${\cal T}(\k,\omega)=\left|G_{\text{C}} (\k,\omega)\right|^2$. In this manner we obtain the finite-momentum probe results shown in Fig.~1 of the main text.

\section{Bound states of three excitons}
\label{sec:trimer}
We now discuss the existence of the vacuum bound three-body triexciton state (trimer) consisting of a spin $\down$ exciton and two spin $\up$ excitons. Assuming that only distinguishable excitons interact and that this occurs via contact interactions (the scenario described in the main text), the problem becomes very similar to few-body problems studied in the context of nuclear physics. In particular, the case of three identical bosons confined to two dimensions was considered as early as 1979~\cite{Bruch1979}.  Here, two trimers exist with binding energies proportional to the dimer (or, in our case, the biexciton) binding energy, $\eb$. The difference in the present case is that rather than having three pairs of bosons that interact, there are only two. The three-body problem is then governed by the implicit equation
\begin{align} 
&-C_{\mathbf{k}}\log \frac{-E+\frac{3}{2}\omega_{{\rm X}\k}}{\eb}
=\frac{4\pi}{m}\sum_{\k'}\frac{\chi(|\k-\k'/2|)\chi(|\k'-\k/2|)}{E-\omega_{{\rm X}\k}-\omega_{{\rm X}\k'}-\omega_{{\rm X},\k+\k'}}C_{\k'},
    \label{eq:2dtrimer}
\end{align}
where $\chi(|\k|)$ is a function that cuts off the integration in the ultraviolet limit. Here, we take $\chi(|\k|)=e^{-k^2/\Lambda_3^2}$, with $\Lambda_3$ being the cutoff momentum. The trimer energy corresponds to the appearance of a pole in the function $C_\k$ for energy $E<-\eb$, and the model presented in the main text corresponds to taking $\Lambda_3\to\infty$. Solving Eq.~\eqref{eq:2dtrimer} in that case, we predict the existence of a single trimer with energy $\varepsilon_T=-2.39\eb$.

\begin{figure}[h]
\centering
\includegraphics[width=.5\columnwidth]{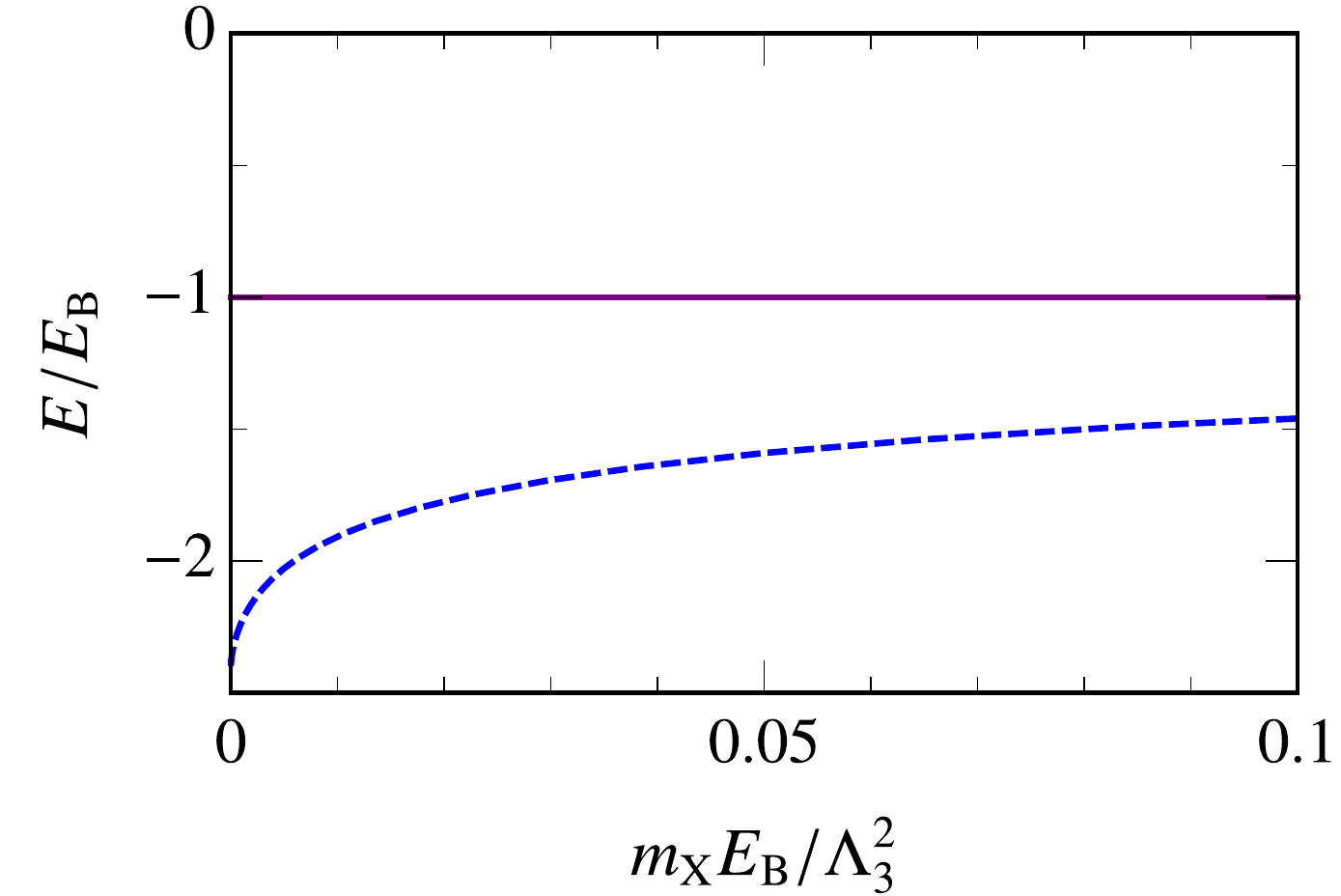}
\caption{Dimer (purple, solid) and trimer (blue, dashed) energies as a function of dimer binding energy in units of the ultraviolet energy scale $\Lambda_3^2/\mx$, see Eq.~\eqref{eq:2dtrimer}.}
\label{fig:trimer}
\end{figure}

The introduction of the ultraviolet cutoff $\Lambda_3$ does not change the two-body problem, but a finite $\Lambda_3$ reduces the strength of the exchange 3-body term. As such, the cutoff allows us to mimic an effective repulsion between identical excitons. In the regime where the length scale associated with repulsion is much smaller than the biexciton size, we find the trimer binding energy shown in Fig.~\ref{fig:trimer}. This explicitly demonstrates that the trimer is a robust feature that should also be expected to exist in the case of a physically realistic repulsion between spin $\up$ excitons.

\end{document}